\documentstyle[epsf]{article}

\begin{document}

\newfont{\bsftitle}{cmssbx10 scaled\magstep2}	
\newfont{\bsfsec}{cmssbx10 scaled\magstep1}	
\newfont{\bsfssec}{cmssbx10}			

\newcommand{\bib}{\noindent
		  \hangindent=4mm
		  \hangafter=1 {}}

\noindent
{\bsftitle 
Measuring long-range dependence
in electricity prices}

\vspace*{.9cm}

\noindent
Rafa{\l} Weron

\vspace*{.4cm}

\noindent
{\small
Hugo Steinhaus Center for Stochastic Methods,\\
Wroc{\l}aw University of Technology, 50-370 Wroc{\l}aw, Poland.}

\vspace*{1cm}

\noindent
{\bf Summary.} The price of electricity is far more volatile than that of other commodities 
normally noted for extreme volatility. The possibility of extreme price movements increases 
the risk of trading in electricity markets. However, underlying the process of price returns 
is a strong mean-reverting mechanism. We study this feature of electricity returns by means 
of Hurst R/S analysis, Detrended Fluctuation Analysis and periodogram regression.

\vspace*{.5cm}


\vspace*{.2cm}

\section*{\bsfsec 1. Introduction}

There exists a strong evidence that price processes of financial assets should not be modeled
by simple random walks (Bouchaud and Potters 1997, Weron and Weron 1998, Mantegna and 
Stanley 1999). These processes seem to be persistent with memory lasting up to a few 
years (Peters 1994) and possess a non-trivial autocorrelation structure (Dacorogna et al. 
1993, Guillaume et al. 1997). 
Recently it has been observed that, contrary to most financial assets, electricity price 
processes are mean-reverting or anti-persistent (Pilipovic 1998, Kaminski 1999, Weron 2000, 
Weron and Przyby{\l}owicz 2000). In this paper we investigate it more thoroughly by means 
of Hurst R/S analysis, Detrended Fluctuation Analysis and periodogram regression.

\section*{\bsfsec 2. Power markets}

The last decade has witnessed radical changes in the structure of electricity markets
world-wide. Prior to the 1980s it was argued convincingly that the electricity industry 
was a natural monopoly and that strong vertical integration was an obvious and
efficient model for the power sector. In the 1990s, technological advances 
suggested that it was possible to operate power generation and retail supply as
competitive market segments (International Chamber of Commerce 1998, Masson 1999).

Changes came slowly at first, reflecting industry concern that competition and system 
security were mutually exclusive. Early experiments -- in Scandinavia and in the UK --
demonstrated clearly that the lights did not go out with the institution of competition.

The deregulation process has recently intensified in Europe and North America, 
where market forces have pushed legislators to begin removing artificial 
barriers that shielded electric utilities from competition. 
Organizations which have been used to long-term fixed price contracts are now 
becoming increasingly exposed to price volatility and, of necessity, are seeking
to hedge and speculatively trade to reduce their exposure to price risk.
However, we have to bear in mind that electricity markets are not anywhere near as
straightforward as financial or even other commodity markets. They have to deal with 
the added complexity of physical substance, which cannot simply be manufactured, 
transported and delivered, at the press of a button.

Unlike other commodities, electricity cannot be stored efficiently. Therefore, a delicate 
balance must be maintained between generation and consumption -- 24 hours a day, 7 days a week, 
52 weeks a year. 
Electric power may be generated from natural gas, coal, oil, nuclear fuel, falling water, 
geothermal steam, alternative resources such as cogeneration, and from renewable resources 
such as wind power, solar energy and biomass. 
Although the principles of generating electricity are simple, generating electricity for 
a country or a state the size of California, both in terms of geographic area and population, 
means a complex balancing process. Naturally, this has big impact on electricity prices and
results in behavior not observed in the financial or even other commodity markets. It is 
thus extremely interesting to investigate the newly established power markets.

\section*{\bsfsec 3. Estimation of long-range dependence}

In economics and finance, long-range dependence has a long history (for a review see 
Baillie and King (1996) and Mandelbrot (1997)) and still is a hot topic of active research 
(Lux 1996, Lobato and Savin 1998, Willinger et al. 1999, Grau-Carles 2000). 
Historical records of economic and financial data typically exhibit nonperiodic cyclical 
patterns that are indicative of the presence of significant long memory. However, the 
statistical investigations that have been performed to test long-range dependence have 
often become a source of major controversies, especially in the case of stock returns. 
The reason for this are the implications that the presence of long memory has on many 
of the paradigms used in modern financial economics (Lo 1991).

Various estimators of long-range dependence have been proposed. In this paper we apply
rescaled range analysis, Detrended Fluctuation Analysis and periodogram regression 
to measure long memory in electricity prices.

\subsection*{\bsfssec 3.1~ R/S analysis}

We begun our investigation with one of the oldest and best-known methods, the so-called
rescaled range or R/S analysis. This method, proposed by Mandelbrot and Wallis (1969) 
and based on previous hydrological analysis of Hurst (1951), allows the calculation of the 
self-similarity parameter $H$, which measures the intensity of long-range dependence 
in a time series. 

The analysis begins with dividing a time series (of returns) of length $L$ into $d$ 
subseries of length $n$. Next for each subseries $m=1,...,d$: 
1$^{\circ}$ find the mean ($E_m$) and standard deviation ($S_m$); 
2$^{\circ}$ normalize the data ($Z_{i,m}$) by subtracting the sample mean 
$X_{i,m}=Z_{i,m}-E_m$ for $i=1,...,n$;
3$^{\circ}$ create a cumulative time series $Y_{i,m}=\sum_{j=1}^i X_{j,m}$ for $i=1,...,n$;
4$^{\circ}$ find the range $R_m = \max \{Y_{1,m},...,Y_{n,m}\} - \min \{Y_{1,m},...,Y_{n,m}\}$;
and 5$^{\circ}$ rescale the range $R_m/S_m$. Finally, calculate the mean value $(R/S)_n$ 
of the rescaled range for all subseries of length $n$.

It can be shown that the R/S statistics asymptotically follows the relation $(R/S)_n \sim cn^H$.
Thus the value of $H$ can be obtained by running a simple linear regression over a sample of
increasing time horizons 
\begin{equation}
\log (R/S)_n = \log c + H \log n.
\end{equation}

Equivalently, we can plot the $(R/S)_n$ statistics against $n$ on a double-logarithmic paper.
If the returns process is white noise then the plot is roughly a straight line with slope 0.5. 
If the process is persistent then the slope is greater than 0.5; if it is anti-persistent then 
the slope is less than 0.5. 

However, it should be noted that for small $n$ there is a significant deviation from 
the 0.5 slope. For this reason the theoretical (i.e. for white noise) values of the R/S 
statistics are usually approximated by 
\begin{equation}\label{eqn-ERS}
{\bf E}(R/S)_n = \left\{
  \begin{array}{lll}
  \frac{n-\frac{1}{2}}{n} \frac{\Gamma(\frac{n-1}{2})}{\sqrt{\pi} \Gamma(\frac{n}{2})}
    \sum\limits_{i=1}^{n-1} \sqrt{\frac{n-i}{i}} & \mbox{for} & n\le 340, \\
  \frac{n-\frac{1}{2}}{n} \frac{1}{\sqrt{n\frac{\pi}{2}}}
    \sum\limits_{i=1}^{n-1} \sqrt{\frac{n-i}{i}} & \mbox{for} & n>340, \\
  \end{array} \right.
\end{equation}
where $\Gamma$ is the Euler gamma function. This formula is a slight modification of 
the formula given by Anis and Lloyd (1976); the $(n-\frac12)/n$ term was added by 
Peters (1994) to improve the performance for very small $n$. 

Formula (\ref{eqn-ERS}) was used as a benchmark in all empirical studies in this paper, 
i.e. the Hurst exponent $H$ was calculated as $0.5$ plus the slope of 
$(R/S)_n - {\bf E}(R/S)_n$. The resulting statistics was denoted by R/S-AL.

A major drawback of the R/S analysis is the fact that no asymptotic distribution 
theory has been derived for the Hurst parameter $H$. The only known results are for the 
rescaled (but not by standard deviation) range $R_m$ itself (Lo 1991). However,
recently Weron (2001) has obtained empirical confidence intervals for the R/S statistics
via a Monte Carlo study. We will use these values in the next Section.

\subsection*{\bsfssec 3.2~ Detrended Fluctuation Analysis}

The second method we used to measure long-range dependence is the Detrended Fluctuation 
Analysis (DFA) proposed by Peng et al. (1994). The advantage of DFA over R/S 
analysis is that it avoids spurious detection of apparent long-range correlation that is an 
artifact of non-stationarity. The method can be summarized as follows. Divide a time series 
(of returns) of length $L$ into $d$ subseries of length $n$. Next for each subseries 
$m=1,...,d$: 
1$^{\circ}$ create a cumulative time series $Y_{i,m}=\sum_{j=1}^i X_{j,m}$ for $i=1,...,n$;
2$^{\circ}$ fit a least squares line $\tilde{Y}_m(x)=a_m x + b_m$ to $\{Y_{1,m},...,Y_{n,m}\}$;
and 3$^{\circ}$ calculate the root mean square fluctuation (i.e. standard deviation) 
of the integrated and detrended time series 
\begin{equation}
F(m) = \sqrt{\frac1n \sum_{i=1}^n (Y_{i,m} - a_m i - b_m)^2}.
\end{equation}
Finally, calculate the mean value of the root mean square fluctuation for all subseries of 
length $n$ 
\begin{equation}
\bar{F}(n)=\frac1d \sum_{m=1}^d F(m).
\end{equation}
Like in the case of R/S analysis, a linear relationship on a double-logarithmic paper 
of $\bar{F}(n)$ against the interval size $n$ indicates the presence of a power-law scaling of 
the form $cn^H$ (Peng et al. 1994, Taqqu et al. 1995). If the returns process is white noise 
then the slope is roughly 0.5. If the process is persistent then the slope is greater than 0.5; 
if it is anti-persistent then the slope is less than 0.5. 

Unfortunately, no asymptotic distribution theory has been derived for the DFA statistics so far. 
However, like for the R/S analysis, Weron (2001) has obtained empirical confidence intervals for 
the DFA statistics via a Monte Carlo study. We will use these values in the next Section.

\subsection*{\bsfssec 3.3~ Periodogram regression}

The third method is a semi-parametric procedure to obtain an estimate of the fractional 
differencing parameter $d$. This technique, proposed by Geweke and Porter-Hudak (1983)
and denoted GPH in the text, is based on observations of the slope of the spectral 
density function of a fractionally integrated series around the angular frequency $\omega = 0$. 
Since the spectral density function of a general fractionally integrated model (eg. FARIMA) 
with differencing parameter $d$ is identical to that of a fractional Gaussian noise with Hurst 
exponent $H=d+0.5$, the GPH method can be used to estimate $H$.

The estimation procedure begins with calculating the periodogram, which is a sample analogue 
of the spectral density. For a vector of observations $\{x_1,...,x_L\}$ the periodogram is 
defined as
\begin{equation}
I_L(\omega_k)=\frac{1}{L} \left|\sum_{t=1}^{L} x_t e^{-2\pi i (t-1) \omega_k} \right|^2,
\end{equation}
where $\omega_k = k/L$, $k=1,...,[L/2]$ and $[x]$ denotes the largest integer less then or 
equal to $x$. The next and final step is to run a simple linear regression 
\begin{equation}\label{gph-reg}
\log\{I_L(\omega_k)\} 
= a - \hat{d} \log\left\{4 \sin^2 (\omega_k/2) \right\} + \epsilon_k,
\end{equation}
at low Fourier frequencies $\omega_k$, $k=1,...,K \le [L/2]$. The least squares estimate 
of the slope yields the differencing parameter $d$ through the relation $d = \hat{d}$, 
hence $H = \hat{d}+0.5$.
A major issue on the application of this method is the choice of $K$. Geweke and Porter-Hudak 
(1983), as well as a number of other authors, recommend choosing $K$ such that 
$K=[L^{0.5}]$, however, other values (eg. $K=[L^{0.45}]$, $[L^{0.2}]\le K\le [L^{0.5}]$) have 
also been suggested.

Periodogram regression is the only of the presented methods, which has known asymptotic 
properties. Inference is based on the asymptotic distribution of the estimate
\begin{equation}\label{gph_var}
\hat{d} \sim N\left(d,\frac{\pi^2}{6 \sum_{k=1}^K (x_t - \bar{x})^2}\right),
\end{equation}
where $x_t = \log\{4 \sin^2 (\omega_k/2)\}$ is the regressor in eq. (\ref{gph-reg}).

\section*{\bsfsec 4. Empirical analysis}

The first analyzed database was obtained from the University of California Energy Institute 
(UCEI). Among other data it contains market clearing prices from the California Power 
Exchange (CalPX) -- a time series containing system prices of electricity for every hour 
since April 1st, 1998, 0:00 until December 31st, 2000, 24:00. Because the series included 
a very strong daily cycle we created a 1006 days long sequence of average daily prices 
and plotted it in Fig. 1. 
The price trajectory suggests that the process does not exhibit a regular annual cycle. 
Indeed, since June 2000, California's electricity market has produced extremely high 
prices and threats of supply shortages.\footnote{The difficulties that have appeared 
are intrinsic to the design of the market, in which demand exhibits virtually no price
responsiveness and supply faces strict production constraints. As yet there is 
no happy end to this story. On January 30th, 2001 the exchange suspended trading because 
it could not comply with FERC's (Federal Energy Regulatory Commission) directive not to 
allow bidding that is inconsistent with the mandated \$150 breakpoint.
Five weeks later, on March 9th, 2001, the California Power Exchange filed for Chapter 11 
protection with the U.S. Bankruptcy Court. This was a serious blow to all protagonists of 
the power market liberalization.} 
We decided to treat this period as an anomaly and remove it from the data when measuring 
long-range dependence. Further analysis was conducted on a 731 days long sequence
of average daily prices covering the period April 1st, 1998 -- March 31st, 2000, i.e. 
two full years.

\begin{figure}[tbp]
\centerline{\epsfxsize=10cm \epsfbox{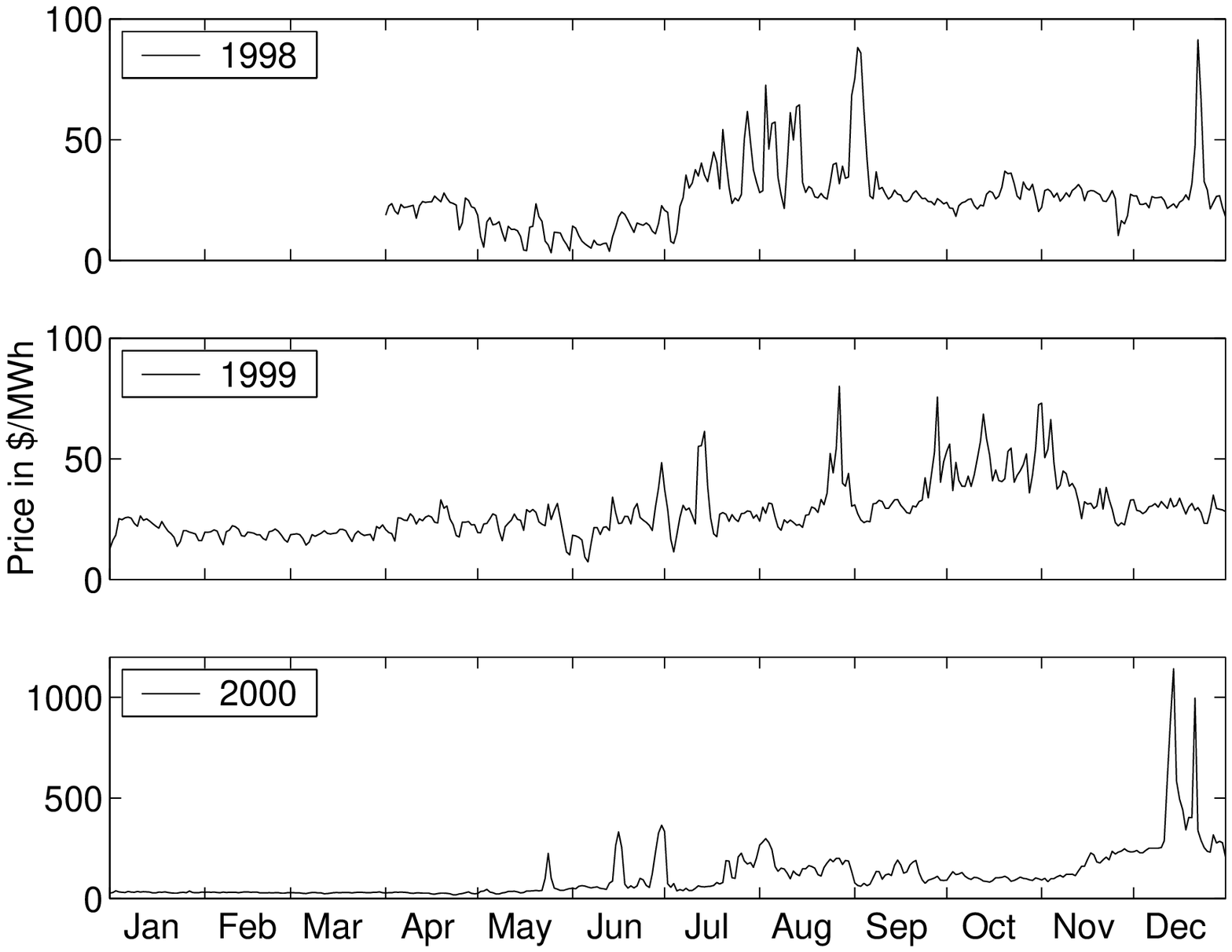}}
{\small
{\bf Fig. 1.} 
CalPX market daily average clearing prices since April 1st, 1998 until December 31st, 2000. 
Note the different scale in the bottom panel.
}
\end{figure}

\begin{figure}[tbp]
\centerline{\epsfxsize=10cm \epsfbox{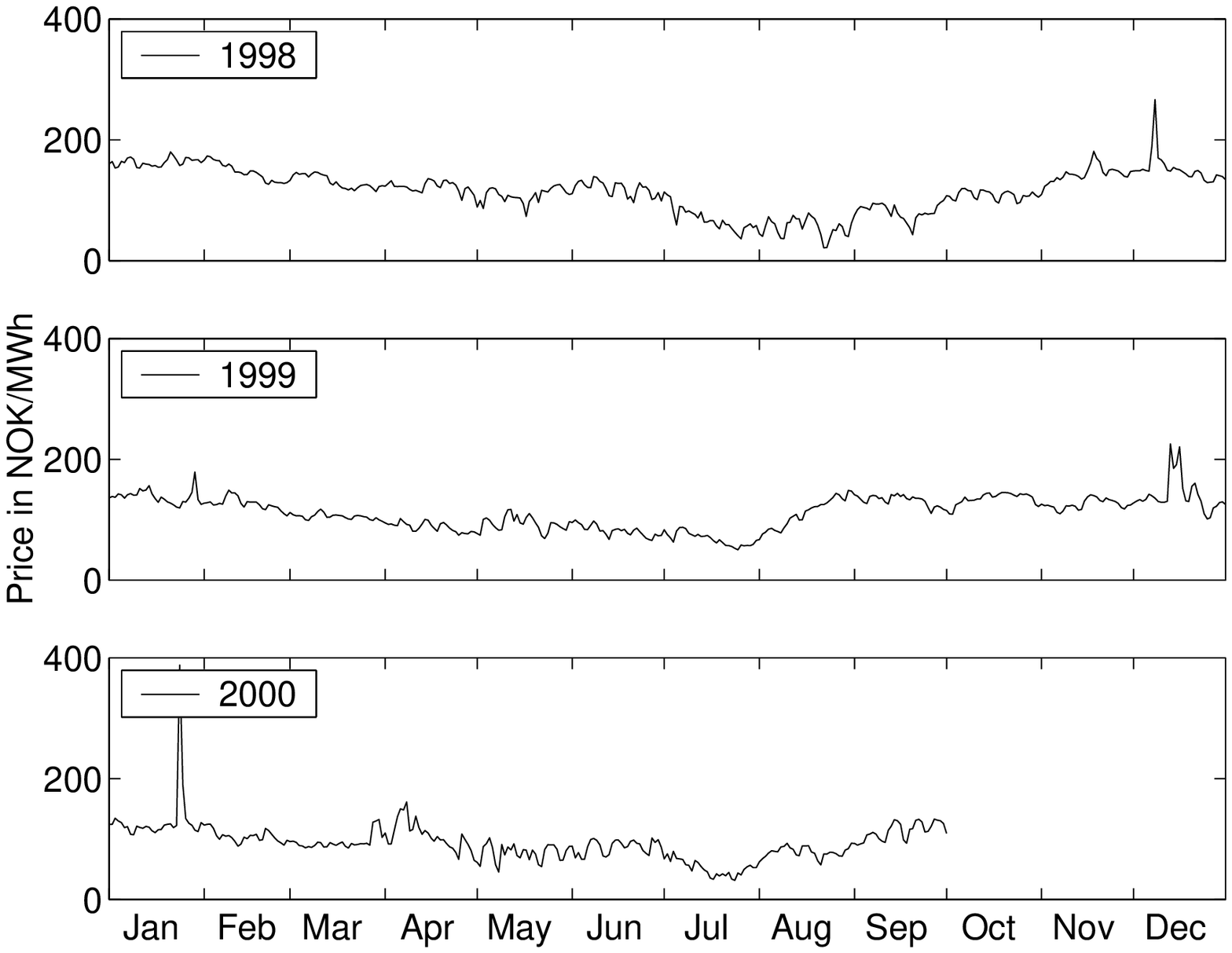}}
{\small
{\bf Fig. 2.} 
Nord Pool market daily average system prices since January 1st, 1998 until September 30th, 2000. 
The annual Scandinavian cycle (low prices in summer, high in winter) can be seen easily. 
}
\end{figure}

All other analyzed time series were kindly provided by Bridge Information Systems. Most of the 
data sets included electricity prices since January 1st, 1998 until September 30th, 2000, 
however, for the analysis we selected data as indicated below:
\begin{itemize}
\item 
\vspace*{-.7em}
a 731 days long sequence of Nord Pool (Nordic Power Exchange) average daily system prices 
(electricity) for the period April 1st, 1998 -- March 31st, 2000 (to be consistent with 
CalPX data); the full series is plotted in Fig. 2;
\item
\vspace*{-.7em}
a 695 days long sequence of daily spot market closing prices (excluding weekends and holidays) 
for firm on-peak power in the Entergy region (Louisiana, Arkansas, Mississippi and East Texas) 
for the period January 2nd, 1998 -- September 29th, 2000;
\item
\vspace*{-.7em}
a 295 days long sequence of Telerate day-ahead U.K. electricity index (Monday through Friday 
only, including holidays) for the period September 1st, 1999 -- September 29th, 2000. 
\end{itemize}
\vspace*{-.7em}

Before we present the results of the empirical analysis observe that R/S and DFA statistics 
require that length $L$ of the data vector has as many divisors as possible. In three 
cases we had to reduce the original number of logarithmic returns in order to increase 
the number of divisors:
for the CalPX and Nord Pool markets we selected the first 728 (out of 730) returns and 
estimated the Hurst exponent using subseries of length $n=52,56,91,104,182$ and 364;
for the Entergy market we selected the first 690 (out of 694) returns and estimated $H$ 
using subseries of length $n=69,115,138,230$ and 345. 
In the case of the U.K. market we only had 294 returns. We decided to use all subseries
of length $n>10$, i.e. 14, 21, 42, 49, 98 and 147, and to calculate only the DFA statistics
(since the rescaled range statistics yields large estimation errors for small $n$).
To keep consistency, periodogram regression (GPH) estimates were obtained for the same data 
sets.
 
Results of the long-range dependence analysis for returns of all four time series are 
reported in Table 1. Hurst exponent $H$ estimates are given together with their 
significance at the (two-sided) 90\%, 95\% or 99\% level. Looking at the table we can 
classify the power markets into two categories:
1$^{\circ}$ those where electricity price processes exhibit a strong mean-reverting 
mechanism and 2$^{\circ}$ those where electricity prices behave almost like Brownian motion. 
The California and Entergy markets fall into the first category, whereas the Scandinavian
market behaves in a more random walk like fashion. Unfortunately, the short length of the 
fourth data set makes the results highly questionable and does not allow us to assign 
the U.K. day-ahead spot market to any category.

\begin{table}[htbp]
{\small
{\bf Table 1.} 
Estimates of the Hurst exponent $H$ for original and deseasonalized data
}
\begin{center}
\begin{tabular}{llll}
\hline
\\[-9pt]
           & & Method & \\
Data       & R/S-AL~~~~~~ & DFA~~~~~~~~~~~~ & GPH~~~~~~~~~~ \\[2pt]
\hline
\\[-9pt]
{\it Original data} & & & \\[2pt]
CalPX      & 0.3473$^*$ & 0.2633$^{***}$ & 0.0667$^{***}$ \\
Nord Pool  & 0.4923 & 0.4148 & 0.1767$^{**}$ \\
Entergy    & 0.2995$^{**}$ & 0.3651$^{**}$ & 0.0218$^{***}$ \\
U.K. spot  & ---    & 0.1330$^{***,{\rm a}}$ & 0.1623$^*$ \\[2pt]
\hline
\\[-9pt]
{\it Deseasonalized data~~~~~~~} & & & \\[2pt]
CalPX      & 0.3259$^*$ & 0.2529$^{***}$ & 0.1336$^{**}$ \\
Nord Pool  & 0.5087 & 0.4872 & 0.3619 \\[2pt]
\hline
\end{tabular}
\end{center}
{\small
$^*$, $^{**}$ and $^{***}$ denote significance at the (two-sided)
90\%, 95\% and 99\% level, respectively. For the R/S-AL and DFA statistics inference 
is based on empirical Monte Carlo results of Weron (2001), whereas for the GPH statistics 
-- on asymptotic distribution of the estimate of $H$.\\
$^{\rm a}$ Due to the small number of data points the DFA statistics for U.K. spot prices
was calculated using subseries of length $n>10$. 
}
\end{table}


To test if these results are an artifact of the seasonality in the electricity price 
process we applied a technique proposed in Weron et al. (2001) to remove the weekly 
and annual cycles in the two longest time series (CalPX and Nord Pool markets). 
The results, which are reported in Table 1, show that mean-reversion is not caused
by seasonality. The estimated Hurst exponents for the California market are almost 
identical to the original ones. In the case of the Nord Pool data the changes are also
not substantial (except for the GPH estimate) and allow to reject long-range dependence.
This random walk like behavior of prices is probably caused by the fact that 
the Scandinavian market is more stable than the U.S. or U.K. markets, with the majority 
of electricity being produced "on-demand" by hydro-storage power plants.

\section*{\bsfsec 5. Conclussions}

Our investigation of the power markets shows that there is strong evidence for 
mean-reversion in the returns series, which -- what is important -- is not an artifact 
of the seasonality in the electricity price process. This feature distinguishes 
electricity markets from the majority of financial or commodity markets, where there is 
no evidence for long-range dependence in the returns themselves.
This situation calls for new models of electricity price dynamics. Simple continuous-time
models were discussed in Weron et al. (2001), but surely more work has to be done in this 
interesting area.

\section*{\bsfsec Acknowledgements}

Many thanks to Gene Stanley for encouragement, to Taisei Kaizoji for an excellent trip 
to Kamakura, to Thomas Lux for stimulating discussions inside the Great Buddha, 
to Joanna Wrzesi{\'n}ska of Bridge Information Systems for providing electricity 
price data and last but not least to Nihon Keizai Shimbun for financial support.

\section*{\bsfsec References}

\bib 
Anis, A.A., Lloyd, E.H. (1976)
The expected value of the adjusted rescaled Hurst range of independent normal summands.
Biometrica 63: 283-298.

\bib 
Baillie, R.T., King M.L., eds. (1996) 
Fractional differencing and long memory processes. 
{\it Special issue of} Journal of Econometrics 73.

\bib 
Bouchaud, J.P., Potters, M. (1997) 
Theory of Financial Risk (in French).
Alea-Saclay, Eyrolles, Paris. {\it English edition}: Cambridge University Press, 2000.

\bib 
Dacorogna, M.M., M\"uller, U.A., Nagler, R.J., et al (1993) 
A geographical model for the daily and weekly seasonal volatility in the FX market.
J. International Money \& Finance 12: 413-438. 

\bib Geweke, J., Porter-Hudak, S. (1983)
The estimation and application of long memory time series models.
J. Time Series Analysis 4: 221-238. 

\bib 
Guillaume, D.M., Dacorogna, M.M., Dave, R.R., et al (1997) 
From the bird's eye to the microscope: A survey of new stylized facts of the intra-daily
foreign exchange markets. 
Finance \& Stochastics 1: 95-129.

\bib 
Hurst, H.E. (1951)
Long-term storage capacity of reservoirs.
Trans. Am. Soc. Civil Engineers 116: 770-808.

\bib 
International Chamber of Commerce (1998) 
Liberalization and privatization of the Energy Sector. 
Paris.

\bib 
Kaminski, V., ed. (1999) 
Managing Energy Price Risk. 
2nd. ed., Risk Books, London. 

\bib 
Lo, A.W. (1991) 
Long-term dependence in stock market prices.
Econometrica 59: 1279-1313.

\bib 
Lobato, I.N., Savin, N.E. (1998)
Real and spurious long-memory properties of stock-market data.
J. Business \& Economic Statistics 16: 261-268.

\bib Lux, T. (1996)
Long-term stochastic dependence in financial prices: evidence from the German stock market.
Applied Economics Letters 3: 701-706.

\bib 
Mandelbrot, B.B. (1997) 
Fractals and Scaling in Finance. 
Springer, Berlin.

\bib
Mandelbrot, B.B., Wallis, J.R. (1969)
Robustness of the rescaled range R/S in the measurement of noncyclic long-run statistical dependence.
Water Resources Res. 5: 967-988.

\bib 
Mantegna, R.N., Stanley, H.E. (1999) 
An Introduction to Econophysics: Correlations and Complexity in Finance.
Cambridge University Press, Cambridge.

\bib 
Masson, G.S. (1999)
Competitive Electricity Markets Around the World: Approaches to Price Risk Management. 
In: Kaminski, V. (ed.) Managing Energy Price Risk. 2nd. ed., Risk Books, London, pp 167-189.

\bib 
Peng, C.-K., Buldyrev, S.V., Havlin, S., et al (1994)
Mosaic organization of DNA nucleotides.
Phys. Rev. E 49: 1685-1689.

\bib 
Peters, E.E. (1994) 
Fractal Market Analysis. 
Wiley, New York.

\bib 
Pilipovic, D. (1998)
Energy Risk: Valuing and Managing Energy Derivatives. 
McGraw-Hill, New York. 

\bib 
Taqqu, M.S., Teverovsky, V., Willinger, W. (1995)
Estimators for long-range dependence: an empirical study.
Fractals 3: 785-788.

\bib 
Weron, A., Weron, R. (1998)
Financial Engineering: Derivatives Pricing, Computer Simulations, Market Statistics (in Polish). 
WNT, Warsaw.

\bib 
Weron, R. (2000)
Energy price risk management.
Physica A 285: 127-134.

\bib 
Weron, R. (2001)
Estimating long range dependence: finite sample properties and confidence intervals.
arXiv: cond-mat/0103510.

\bib 
Weron, R., Przyby{\l}owicz, B. (2000)
Hurst analysis of electricity price dynamics.
Physica A 283: 462-468.

\bib 
Weron, R., Koz{\l}owska, B., Nowicka-Zagrajek, J. (2001)
Modeling electricity loads in California: a continuous time approach.
In: Proceedings of the NATO ARW on Application of Physics in Economic Modelling, Prague 2001,
Physica A, in press. See also: arXiv: cond-mat/0103257.

\bib
Willinger, W., Taqqu, M.S., Teverovsky, V. (1999)
Stock market prices and long-range dependence.
Finance \& Stochastics 3: 1-13.

\end{document}